# Industrial-Strength Verification of Solid State Interlocking Programs

**Alexei Iliasov**

The Formal Route

**Dominic Taylor**

Systra Scott Lister

**Linas Laibinis**

Vilnius University

**Alexander Romanovsky[1]**

The Formal Route and Newcastle University

**Abstract** *The increasing complexity of modern interlocking poses a major challenge to ensuring railway safety. This calls for application of formal methods for assurance and verification of their safety. We have developed an industry-strength toolset, called SafeCap, for formal verification of interlockings. Our aim was to overcome the main barriers in deploying formal methods in industry. The approach proposed verifies interlocking data developed by signalling engineers in the ways they are designed by industry. It ensures fully-automated verification of safety properties using the state of the art techniques (automated theorem provers and solvers), and provides diagnostics in terms of the notations used by engineers. In the last two years SafeCap has been successfully used to verify 26 real-world mainline interlockings, developed by different suppliers and design offices. SafeCap is currently used in an advisory capacity, supplementing manual checking and testing processes by providing an additional level of verification and enabling earlier identification of errors. We are now developing a safety case to support its use as an alternative to some of these activities.*

---

[1] Contact author's email: alexander.romanovsky@formal-route.com

# 1 Railway Signalling

Effective signalling is essential to the safe and efficient operation of any railway network. Whether by mechanical semaphores, coloured lights or electronic messages, signalling allows trains to move only when it is safe for them to do so. Signalling locks moveable infrastructure, such as the points that form railway junctions, before trains travel over it. Furthermore, signalling often actively prevents trains travelling further or faster than is safe.

There are two main safety principles shared by all signalling systems:

- A schema must be free from collisions. A collision happens when a train occupies the same physical space as another train or (at a level crossing) a road vehicle. Signalling systems uphold this principle through the use of signalling routes, and block sections.
- A schema must be free from derailments. A derailment may happen when a set of points moves underneath a train. To avoid this, a point must be positively confirmed to be locked in position before a train may travel over it and held in that position as a train does so.

At the heart of any signalling system there are one or more interlockings. These devices constrain authorisation of train movements as well as movements of the infrastructure to prevent unsafe situations arising. One of the earliest forms of computer-based interlocking was the Solid State Interlocking (SSI) (Cribbens 1987), developed in the UK in the 1980s through an agreement between British Rail and two signalling supply companies, Westinghouse and GEC General Signal. SSI is the predominant technology used for computer-based interlockings on UK mainline railways. It also has applications overseas, including in India, Australia, New Zealand, France and Belgium. Running on bespoke hardware, SSI software consists of a core application (common to all signalling schemes) and site-specific geographic data. SSI GDL (Geographic Data Language) data configures a signalling area by defining site specific rules, concerning the signalling equipment as well as internal latches and timers that the interlocking must obey. Despite being referred to as data, an SSI GDL configuration resembles a program in a procedural programming language and, as such, is referred to as a program in this paper. Such a configuration is iteratively executed in a typical loop controlling the signalling equipment.

The increasing complexity of modern digital interlockings, both in terms of their geographical coverage and that of their functionality, poses a major challenge to ensuring railway safety. This calls for application of rigorous methods, in particular formal methods, for assurance and verification of safety and other crucial properties of such systems. Even though formal methods have been successfully used in the railway domain (e.g. (Badeau and Amelot 2005)), their industry application is scarce, typically focusing on relatively simple interlockings used in metros and other urban lines. In spite of a large body of academic studies addressing issues of formal verification of railway signalling systems, they usually remain an academic exercise due to a prohibitive cost of initial investment into their industrial deployment as well as the inherent limitations of the formal techniques used (such as the state explosion for model checking techniques or substantial manual efforts for applying provers). There are a number of reasons for this;

- signalling engineers need to learn mathematical notations and formal reasoning to effectively apply them;
- many verification techniques and the supporting tools proposed cannot be applied to analysing real modern interlockings due to their poor scalability;
- companies need to drastically change the existing development processes in order to use them;
- the development of formal techniques and tools in academia is seldom driven by the chief aim of deploying them in industry.

The paper first introduces a novel tool-based approach that addresses the above issues by

- verifying the signalling programs and layouts (schemas) designed by signalling engineers in the ways they are developed by industry,
- ensuring fully-automated and scalable verification of safety properties using the state of the art verification techniques (in particular, automated theorem provers and solvers), and
- providing diagnostics in terms of the notations used by the engineers.

All together, this ensures that the developed method and tool can be easily deployed to augment the existing industrial processes of developing complex *mainline* interlockings in order to provide extra guarantees of railway safety.

The remaining part of the paper discusses the proposed verification method, its industrial deployment and application for safety verification of a substantial number of live projects conducted by major signalling companies in the UK, as well as our ongoing work towards developing a safety case to allow the approach to be deployed as a (partial) replacement for the manual checks and testing/simulation widely used now for safety assurance by the railway industry.

Our earlier paper (Iliasov et al 2018) discusses a prototype (experimental) version of the method and the tool, provides additional information on how the SSI verification is conducted and discusses a small case study. Since this earlier paper was written the tool has been substantially reworked and enhanced to improve its usability, scalability, reliability, and the quality of reporting, and to extend the verification coverage. During this period the tool has been demonstrated, validated, and proven in an operational environment, and approved by the cross-industry SSI-Applications Group led by Network Rail for use in UK signalling projects for automated railway signalling verification. All this has allowed us to deploy the approach on multiple commercial railway signalling projects in the UK.

## 2 SafeCap for Solid State Interlocking

The SafeCap platform (Iliasov et al 2013, Iliasov et al 2014) is a general open ex-tandable Eclipse-based toolkit (Des Rivieres and Wiegand 2004) for modelling rail-way capacity and verifying railway network safety developed in a number of public projects led by Newcastle University. The platform's main purposes have been to support academic research and to help in exploring how new ideas could be effi-ciently used by the railway industry. It allows the users to design stations and junc-tions relying on the provided domain specific language (SafeCap DSL - (Iliasov and Romanovsky 2012)) and to check their safety properties, simulate train runs as well as evaluate potential improvements of railway capacity by using a combination of theorem proving, SMT solving and model checking.

The SafeCap DSL allows the designers to rigorously and unambiguously define a model of the given railway network (e.g., stations or junctions) by providing a formal, graph-oriented way of capturing railway schemas and some aspects of sig-nalling. Various concepts of a railway schema such as signals and signalling solu-tions, speed limits, stopping points and so on can be incorporated via DSL extension plug-ins. Such plug-ins introduce new data (as custom annotations) and the support-ing logic (as additional logical constraints or relationships). Such a tool architecture allows us not to commit to any regional technology and thus to offer a broadly sim-ilar approach for a range of legacy and current technologies.

The SafeCap verification and proof back-ends enable automated reasoning about static and dynamic properties of railways or their signalling data. In the course of platform development, it has been substantially extended by adding new simulators, state-of-the-art solvers and provers, as well as the support for importing the existing designs in a wide range of signalling frameworks supported by industry.

In the last 5 years, initially using the SafeCap platform as the experimental pro-totype, a team of computer scientists and signalling engineers has been working on developing a targeted SafeCap-based industry-strength toolset for formal verifica-tion in the course of the SSI development projects. Our aim from the outset was to overcome the typical barriers in deploying formal methods in industry. In the course of this work we have removed the SafeCap functionalities that are not required for SSI verification (such as animation, simulation, and capacity measurement), sim-plified the tool architecture, and replaced all its solvers and provers with a dedicated inference-based symbolic prover that outperforms all known state-of-the-art prov-ers when used on the railway models that our SSI-targeting tool needs to verify. The core requirements set for this work were that the method and the tool:

- use the existing industrial signalling notations for the inputs and for reporting the results of verification;
- fully hide formal methods from the engineers that use them;
- provide fully automated verification;
- are scalable to allow full safety verification of any existing UK interlocking within a few minutes.

As we explained earlier an SSI GDL configuration resembles a program in a procedural programming language. An SSI system is a continuously running control system that executes a global control loop composed of three stages:

1. polling of inputs (the current states of train detection, points, signals and so on);
2. computation of necessary responses, as well as,
3. formation and transmission of equipment control commands.

The input and output stages are generic and their safety argument is provided once for a particular underlying hardware implementation. SafeCap verification of system safety is concerned with the middle stage – the response computation; the safety of other stages is assured through the safety cases underpinning the generic interlocking hardware and the site-specific wiring connecting that hardware to trackside equipment. This stage is unique to every geographic area and explicitly refers to the equipment drawn on the area *scheme plan* – a diagrammatic depiction of a railway.

The central question in the verification of signalling correctness is what constitutes a safe signalling design. Certain basic principles are universally accepted, for instance, the absence of train collisions and derailment. However, it is almost hopeless to verify the absence of such hazards in the strictest possible sense, not least because there are many real-world limitations:

- trains are driven in accordance with signals, but signals can only convey very limited information, which is only readable for a short distance on the approach to each signal;
- trains have different braking characteristics and must all be given sufficient warning of the need to stop before a red signal;
- drivers occasionally misjudge braking and pass red signals;
- there are some situations that cannot be fully protected by a signalling system, such as allowing trains to couple together or share a platform, for which the basic principles are upheld through human competence and operational procedures.

Instead, correctness is established not against the basic principles but rather against *signalling principles* derived from the basic principles and designed to enable railway operation with an acceptable level of risk and failures: the conditions under which trains can be authorised to move by signals; the indications ('aspects') displayed by those signals and their meanings; when points can move; provision of 'overlaps' beyond red signals in which a train can stop safely if the driver misjudges braking; etc.. Such signalling principles are carefully designed by domain experts and documented in standards such as (Network Rail 2015). However, they can vary between regions and do change over time. At the time of writing this paper the SSI SafeCap verifies 55 signalling principles.

For the purposes of verification, each signalling principle to be ensured is rendered as an *inductive safety invariant* – a system property that must hold when a system boots up and must be maintained (or, equivalently, be re-established) after any state update.

One example of a SafeCap safety invariant is an invariant which checks that whenever the commanded position of a set of points is changed (triggering the points to move) all train detection sections over the points are proved clear:

```
forall p:Node
 point_c(p) != "point_c'p"(p) =>
  "pointcleartracks"["Node.base"~[{"Node.base"(p)}]]
          /\ track_o == {}
```

As with all safety invariants, before this formal notation is developed we define and agree on the semi-formal formulation of the invariant:

```
[for]
 every point commanded to a new state
[it holds that]
 all point tracks are proved clear
```

Another example is an invariant that checks that on setting a route, all the route sub routes are locked:

```
forall r:Route
 r:cover(route_s \ "route_s'p") =>
   (forall sr:SubRoute
    sr : "route:subrouteset"[{r}] /\ "SubRoute.ixl" =>
           sr : subroute_l
         )
```

This is the semi-formal notation of the invariant:

```
[for]
 every route being set
[it holds that]
 all the interlocking-contained sub routes of
                       the route are locked
```

Verification of an inductive safety invariant is understood as the problem of checking that any safety invariant is respected by every state update. Technically this is done be generating conjectures (also called proof obligations – POs) of the form "*if an invariant holds in a previous state and a certain state update happens, is it true that the invariant holds for the new state?*". Formally, a conjecture is represented as a logical sequent consisting of a number of hypotheses (*H*) and a goal (*G*), denoted as *H |- G*.

The number of such conjectures is $m * n$, where $m$ is the number of safety invariants (circa 50) and $n$ is the number of possible state updates (typically around 10,000). This is a small number when contrasted against the number of potentially reachable states (circa $2^{2,000}$). The complexity measured in the number of conjectures grows linearly with the system size. In practice, however, it is possible to automatically prove or disprove the vast majority of conjectures with only a tiny number (less than 1 in 10,000 for recent results) of provable conjectures failing to prove and resulting in a false positive. Achieving such a level of proof automation is not easy and requires a number of customised techniques developed and integrated into SafeCap.

The major steps of SafeCap verification process are depicted in the diagram in Figure 1. There are four major stages: preparation of input (steps 0-3), translation and symbolic execution of the signalling logic (steps 4-5), formal verification (steps 6-8), and report generation (step 9). All but the very first stage are automated. In the first stage, input preparation turns an electronic image of a scheme plan into a mathematical model. In the next stage, the obtained model is translated into a large number of individually simple state transitions. In the third stage, formal verification is employed to check that these state transitions are safe. In the last stage, the engineer report is generated to present findings of safety violations in the form understandable to engineers.

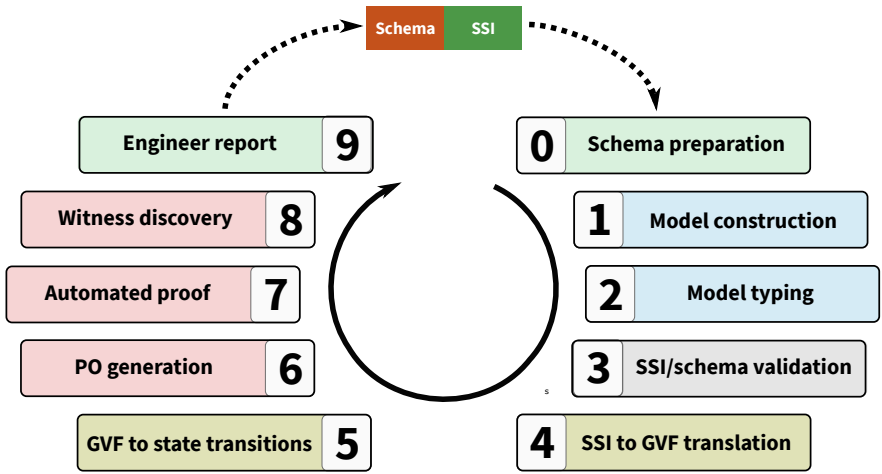

**Fig. 1.** SafeCap verification process

The associated diagram in Figure 2 depicts the overall data flow of the approach and cross-links it with the steps in Figure 1. The corresponding data transformation steps (arrows) are annotated by numbers referring to the respectively numbered steps of Figure 1. Intuitively, the left branch focuses more on the system statics (railway scheme, its various constituent elements and their relationships), while the right one deals with the system dynamics related to SSI signalling. The middle branch covers industrial standards, formulating safety principles relevant to the scope and operation level of SSI and then deriving formal safety invariants. Note that the solid arrows in Figure 2 depict automated actions by the SSI SafeCap tool, while dashed ones are manual activities.

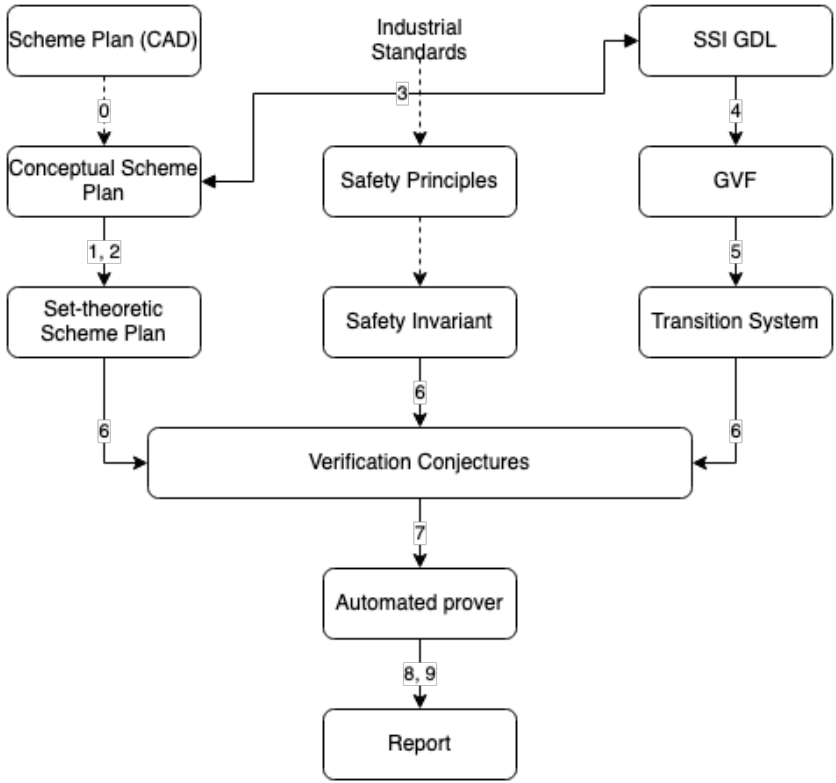

**Fig. 2.** Data flow diagram

In general, each subsequent layer downwards adds more rigour and formality. The data representations of the third layer (i.e., set theoretical scheme plan, safety invariant, and transition system) are based on the same underlying mathematical language and can be considered as parts of the overall formal system model ready to be verified.

At the top level, the inputs defining a particular area are a relevant scheme plan and the associated SSI GDL source code. They both are translated into the corresponding mathematical models in two steps. The intermediate representations, a conceptual scheme plan and GVF (Generic Verification Framework), are needed for practical reasons, mainly, to allow us to deal with idiosyncrasies in a particular input notation, common in railway scheme plans, and to enable a generic verification approach relying on symbolic execution.

The two formal models derived from the scheme plan and SSI GDL are interrelated using the derived safety invariants, which leads to generation of necessary verification conjectures. Any violations of the safety invariants are reported as findings in the final report. All model transformations during steps 0-9 in Figure 1 are traceable in both directions via meta-references created during the tool operation to allow the tool to create the final report describing the findings in terms of the initial inputs: for each violation it includes its main characteristics as well the relevant part of the SSI code and the schema plan fragment (several extracts of the verification reports can be found in (Iliasov et al 2018, Taylor et al 2019).

# 3 Industrial experience

In the last two years SafeCap has been extensively used to automatically verify the compliance of real-world SSI GDL data with safety properties. 26 different UK interlockings, developed by several different suppliers and multiple design offices, have been verified. Six of these interlockings were analysed as trial applications of SafeCap, demonstrating that the automated approach could consistently find known errors in the data (both deliberately seeded and unintentional in non-in-service data sets) in far less time than it would take to check or test manually (minutes rather than weeks). The trial applications also showed SafeCap's value in highlighting vulnerabilities in data, where safety depended on the complex interaction of different signalling principles applying to different sections of a railway layout in ways that would not be immediately apparent to a manual designer or checker charged with modifying that data. The remaining 20 were commercial applications of SafeCap in live signalling projects fulfilling industry requirements for automated verification.

Safety property violations, reported by SafeCap in these practical applications, fell into four categories:

- errors – straightforward errors that needed to be corrected;
- vulnerabilities – where the combination of circumstances under which a reported violation occurs can, through other properties, be shown impossible in practice, though it could unintentionally be made possible through modifications to seemingly unrelated data;
- intentional violations – violations of specific properties in specific locations for operational reasons, for example to allow trains to shunt backwards and forwards in a siding without signaller involvement or to allow a train to couple to the front of another train;
- false positives – reported violations where it can be shown that none exists, resulting from limitations of the safety properties as explained below.

Practical experience of SafeCap with real-world data has led to refinements of safety properties to reduce false positives by reflecting the manner in which SSI GDL data is constructed. To make efficient use of once scarce processing power, signalling interlockings do not explicitly test signalling principles in every instance they apply. Instead, they test specific instances and infer compliance in other instances through other signalling principles. Constructing a logical proof of compliance with one signalling principle can require safety properties that explicitly exclude cases that violate other signalling principles and additional safety properties to prove those other principles.

For example, where multiple set of points that are wired to always move together, the interlocking may only test that the last section of route (known as a *sub route*) locking any of the points is free. The fact that other the sub route(s) locking the points is also free is inferred, because it is locked at the same time as, and freed sequentially before, the sub route being tested. This is illustrated in Figure 3.

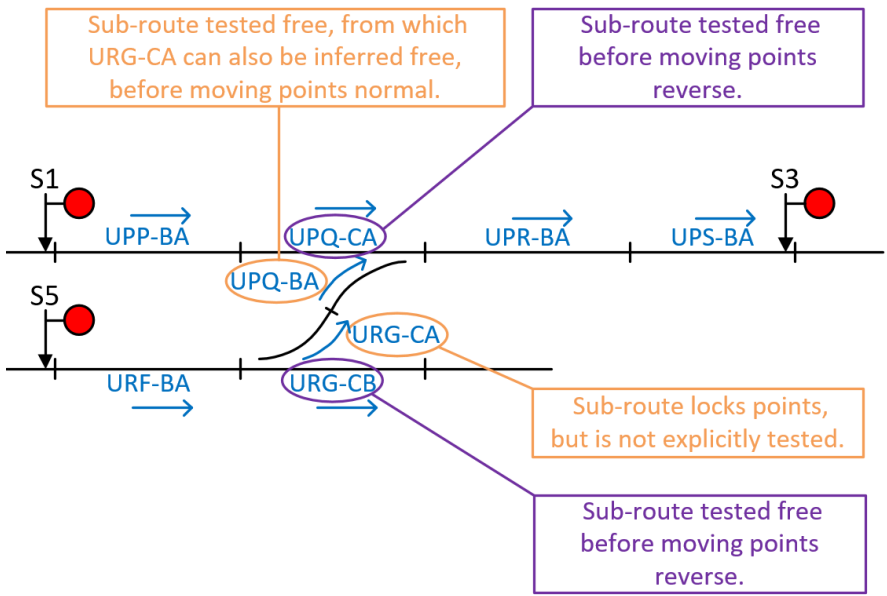

**Fig. 3**. Sub route locking of points

Sometimes the interactions between signalling principles can be even more complex, involving interactions between points and overlaps in apparently unrelated parts of the layout. Whilst such cases can still be proven to be safe by SafeCap, they are unlikely to be obvious to anyone modifying the data and hence represent a safety vulnerability to which SafeCap can alert designers.

Even though the fully automated SafeCap verification does not require any manual proofs, is still accompanied by some manual efforts from the formal method expert and the experienced signalling engineer. In some situations, the former needs to adjust the proof tactics and properties for a specific project, the latter is involved in adjusting the properties and in the interpretation of the automated verification results. The railway expert leads the production of the final report including the categorisation of the violations found and the selection of the counter examples. Some manual efforts are often required to deal with differences in the way the SSI programs and the schemas are produced by different engineering teams and with the mismatches between the SSI program and the schema (typically, naming or naming convention mismatches) in a given project.

Our current aim is to substantially reduce these manual efforts. For example, relying on our significant experience in applying the tool we have already developed a limited set of possible proof tactics, from which we could quickly select the most suitable one when necessary.

The development or re-development of the railway network, including signalling, is naturally structured into projects. Each project constitutes a substantial geographical area (such as a large UK station), controlled by 2-4 interlockings (or virtual interlockings [2]) each of which has an SSI program located on one interlocking controller unit. This design practice restricts the complexity of individual SSI programs; however, our experience shows that it can still vary a lot depending on the way the individual SSI program is written.

**Table 1**. Statistics about verification in three projects

| Name | Routes | Points | Signals | Safety invariants | State transitions | Time, seconds | RAM memory, peak |
|------|--------|--------|---------|-------------------|-------------------|---------------|------------------|
| X | 140 | 48 | 83 | 55 | 11078 | 37 | 42Gb |
| Y | 64 | 29 | 92 | 55 | 4462 | 3 | 4Gb |
| Z | 190 | 64 | 84 | 55 | 112560 | 265 | 67Gb |

Table 1 shows the statistics about the verification conducted in the last three recently completed projects run on a professional PC with 16 cores. The Z project was one of the projects we completed with the longest verification run. We note here that the scheme elements are not a good predictor of verification complexity; instead complexity, especially its upper boundary, is better explained by the maximum cyclomatic code complexity of the verified SSI program. In practice this is often defined by cascading swinging overlaps that tend to require deeply nested subroutine calls. The state space of these SSI projects is extremely large as they have in average 1.5-2.5K Boolean variables and 100-200 Integer variables, but this is irrelevant to the type of verification we conduct.

---

[2] A virtual interlocking is a specific interlocking application running on a shared hardware platform, often with other virtual interlockings. It differs from a traditional electronic interlocking in which there was a one-to-one mapping between interlocking applications and the hardware they ran on.

As part of our work, we have extended the toolset by importing several proprietary schema notations used by our industrial partners and successfully experimented with importing schemas in RailML [3] and SDEF [4].

During our work on the live projects, we continue the improvements of the toolset. First, we have extended the number of the safety invariants to improve the verification coverage. Secondly, we have improved the quality of reporting by reducing the number of false positives. Substantial efforts have been dedicated to improving the quality of the safety invariants with the aim of reducing the verification time and improving the diagnostics of the property violations. During this work, while dealing with a number of interlocking datasets designed by different companies and their different offices, we have developed a library of proof tactics that can improve the proof automation. We continue putting substantial efforts into improving the scalability of the tool to make sure that, even with the extended number of the safety invariants, it takes no longer than 5-6 minutes to verify the safety of any UK signalling interlocking. Finally, we have greatly improved the usability of the native SafeCap schema editor as, in our experience, the vast majority of railway schemas in the UK are available only as PDF images or CAD files from which signalling information cannot readily be gleaned.

Initial commercial applications of SafeCap have been in response to a new UK industry requirement for automated verification, which provides additional mitigation of the risk of error in safety-critical SSI GDL data. However, SafeCap has the potential to offer much greater benefit if used earlier in the design process. Doing so would not only meet the industry requirement for automated verification, but could also enable earlier identification of errors in the data production process, thereby avoiding expensive and time-consuming rework cycles. We are currently working with industry partners on determining the optimal phases for SafeCap analysis within the existing data production process.

## 4 Building a safety case

SafeCap is currently used in an advisory capacity, supplementing manual checking and testing processes by providing an additional level of verification and enabling earlier identification of errors. Whilst this brings significant benefits, greater value could be realised by using SafeCap as an alternative to these costly and time-consuming manual processes. However, to do this it is necessary to demonstrate that it is safe to use SafeCap in this manner.

---

[3] Railway Markup Language. RailML.org. The RailML Standard. Data exchange v.3.1. The RailML.org Initiative. https://www.railml.org/en/introduction.html Accessed 26 August, 2021

[4] Signalling Data Exchange Format. Network Rail Signalling Innovations Group (NR SIG). SNIP – 132890 System Data Exchange Format (SDEF). Version 7.2 Design. 2014 https://www.sparkrail.org/Lists/Records/DispForm.aspx?ID=24659 Accessed 26 August, 2021

Specifically it is necessary to demonstrate that each hazard and associated risk arising from the use of SafeCap in a signalling data preparation process is controlled to an acceptable level. The stringency of the acceptability criteria increases with the dependency placed on SafeCap within the data production process. The current use of SafeCap, as an advisory tool, places minimal dependency on it as proven manual processes remain in place. Greater dependency is placed on it if it is used as an alternative to manual checking or testing. Even greater dependency is placed on it if used as an alternative both manual checking and testing. An iterative approach is being taken to safety case development whereby the case will be progressively extended in stages, as the tool itself develops and matures, to enable greater dependency to be placed on it. The decision to use SafeCap as an alternative to specific manual checking and / or testing processes will be made be made by signalling system suppliers developing data using these processes, informed by a compelling safety case that safety risks are being suitably managed.

The safety case is being developed in accordance with the process outlined in the European Common Safety Method for Risk Evaluation and Assessment (CSM-REA) (European Parliament and Council 2015). The first step of this process is to assess whether a change – in this case the use of SafeCap in data production processes – is deemed 'significant' from a safety management perspective. The authors' assessment is that current use of SafeCap in an advisory capacity is not a 'significant' change, as SafeCap merely supplements existing processes. However, replacing established manual checks / tests by SafeCap verification is believed to be significant and hence necessitating the application of the full CSM-REA process. This is because of the novelty of the SafeCap approach, the complexity of the change and the credible worst-case scenario failure consequence that a serious error in signalling data goes undetected (a 'false negative').

Having established the need to apply CSM-REA, the next step is to define the system for which the safety case is being produced. The System Definition is required to cover the system objectives, system functions and elements, system boundary, physical and functional interfaces, system environment, existing safety measures and assumptions. The safety case, and associated System Definition, for the SafeCap tool, is being developed independently of the properties that SafeCap verifies; the correctness of the safety properties is demonstrated through their traceability to established signalling principles. This enables development of the tool and its associated safety case to be de-coupled from the safety properties.

As the SafeCap tool continues to develop and the dependency placed on it increases, the system description will need updates to reflect the latest objective and construction of SafeCap. Similarly, as safety properties are refined or expanded or new sets of safety properties are produced to cover different technologies / signalling principles so traceability to established signalling principles for specific railways can be undertaken independently of the tool.

Safety hazards (for the SafeCap tool) are identified at the system boundary, as documented in the System Description, and recorded in a SafeCap hazard log. Principle among these are the interpretation of SSI GDL and reporting of safety property violations; failings in either could lead to serious error in signalling data going undetected. Various mitigation measures are being developed to reduce the risk associated with these, and other hazards to a level assessed to be either broadly acceptable or as low as reasonably practicable (ALARP).

In the case of safety properties, the safety argument hinges not on explicit risk assessment as for the tool, but on demonstrating that the safety principles embody specific signalling principles. This approach is referred in CSM-REA as risk acceptance by application of code of practice, the codes of practice in question being the standards in which signalling principles are specified. The argument for the correctness and completeness of these principles stems from the extensive expert review, risk assessment and real-world operational experience, over more than two centuries of railway history that has led to them in their current form. As explained earlier, attempting to improve on this through formal demonstration of the absence of hazards is almost hopeless due to the many real-world limitations.

There is, however, a need to demonstrate that these signalling principles have been correctly encapsulated in formal notation. At a basic level, this is achieved through requirements tracing: identifying the specific safety property(ies) that implement a safety property as expressed in specific clauses in railway standards. However, as explained in Section 3, implementation of one signalling principle can depend on others being upheld.  In such cases, goal structured notation is used to demonstrate how the goal of demonstrating compliance with a signalling principle is upheld through lower level safety properties, as in the example in Figure 4.

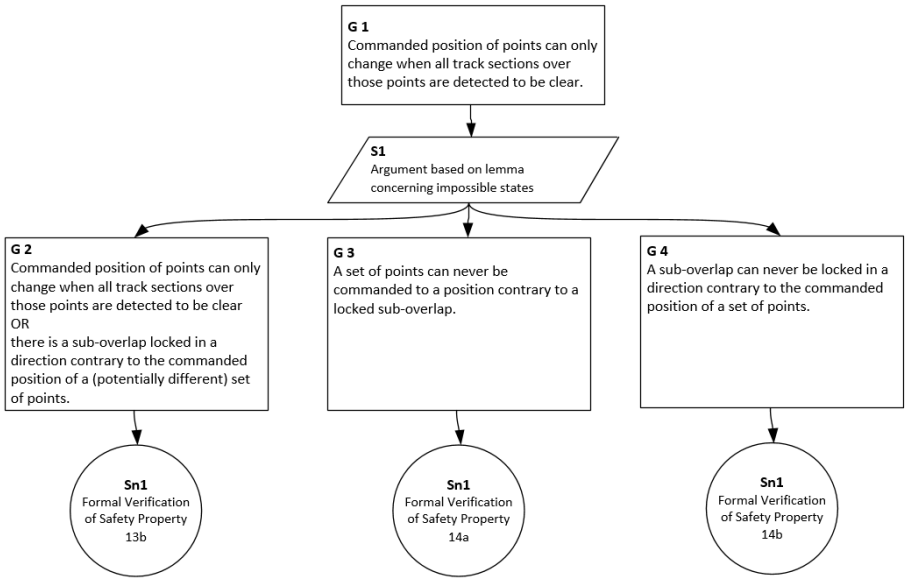

Fig. 4. Goal Structure Notation for example signalling principle

The safety argument also requires application constraints that must be adhered to for the traceability to be valid:

- trackside signalling equipment must be installed in accordance with the signalling plans provided for SafeCap verification;
- interlocking hardware inputs and outputs must be wired consistently with the functions assigned to them in data;
- points that are shown as always moving together on signalling plans must be wired to do so in the real-world;
- etc.

# 5 Conclusions

The paper reports on our successful and substantial deployment of formal methods in the railway industry. The SSI SafeCap tool developed for the verification of the SSI signalling meets the ambitious requirements we set when we started this work (Section 3): the tool uses the signalling and schema notations used by industry, its application does not require any knowledge of formal methods from signalling engineers, the verification is fully automated and hidden from the engineers, the tool is scalable and capable of verifying any mainline signalling within minutes. The tool has been deployed in industry: it is being applied in a number of live signalling projects now.

The research work underlying the development on this tool has been solely driven by the aim of achieving its successful industrial deployment. It has been our principal position from the outset that only with this approach we achieve real deployment of formal methods.

Our future plans focus on improving and extending the safety properties, building a safety case for the tool through which to gain its acceptance as an alternative to manual checking / testing and improving the quality of diagnostics. In the longer term we plan to use the generality and openness of SafeCap to extend the tool with verification of other signalling representations used in metro and mainline signalling in the UK and overseas.